\documentclass[intlimits,twoside,a4paper]{article}

\usepackage[eqsecnum]{cmpj3}

\usepackage[cp1251]{inputenc}
\usepackage{bm}

\usepackage{float}

\newcommand{\ii}{\ri}

\issue{2017}{20}{4}{43705}
\doinumber{10.5488/CMP.20.43705}

\title[On the DOS of graphene]{On the density of states of graphene in the nearest-neighbor approximation}

\author{V.O. Ananyev, M.I. Ovchynnikov}
\address{Faculty of Physics, Taras Shevchenko National Kyiv University, \\6 Academician Glushkov Ave., 03680 Kyiv, Ukraine}

\date{Received June 8, 2017, in final form July 21, 2017}

\begin{document}

\maketitle

\begin{abstract}
We propose an alternative analytical expression for the density of states of a clean graphene in the nearest-neighbor approximation. In contrast to the previously known expression, it can be written as a single formula valid for the whole energy range. The correspondence with the previously known expression is shown and the limiting cases are analyzed.
\keywords graphene, density of states, Van Hove singularity
\pacs 71.20.Tx, 73.22.Pr
\end{abstract}

\section{Introduction}
A theoretical background of the electronic applications of graphene
is based on the knowledge of its band structure. The vast majority of theoretical approaches exploit the fact that the low-energy quasiparticle excitations in graphene have a linear Dirac-like spectrum up to the energies of the order of $0.5$~eV. Widely used simple analytical expressions for the density of states (DOS), optical conductivity, etc.,  ultimately originate from this spectrum \cite{Gusynin2007IJMPB,Neto2009RMP,Katsnelson.book}.

There are very few analytical results describing electronic properties of graphene beyond the continuum linear approximation. Among them there is an expression for the DOS provided by Hobson and Nierenberg in 1953 \cite{Hobson1953PRev} without derivation. It appears  in  various forms in the modern literature (see e.g., review \cite{Neto2009RMP} and reference~\cite{Rammal1985JdPhys}). In particular, in \cite{Katsnelson.book} the DOS per unit cell and one spin component reads
\begin{equation}
D(E) = \left\{
\begin{array}{cc}
\frac{ |\varepsilon|}{t \piup^2} \frac{1}{\sqrt{ F(|\varepsilon|)}}
\mathbb{K}  \left( \frac{|\varepsilon|}{ F(|\varepsilon|)} \right),   & 0 \leqslant |\varepsilon| \leqslant 1, \\
\frac{ |\varepsilon|}{t \piup^2} \frac{1}{\sqrt{ |\varepsilon|}}
\mathbb{K} \left( \frac{F(|\varepsilon|)}{|\varepsilon|} \right), & 1 \leqslant |\varepsilon| \leqslant 3,
\end{array}
\right.\label{DOS-conventional}
\end{equation}
where the energy $\varepsilon = E/t$ is measured in  units of the nearest-neighbor hopping energy $t \approx 3$~eV, the function $g(x)$ is given by
\begin{equation}
F(x) = \frac{(1+x)^2}{4} - \frac{(x^2-1)^2}{16} = \frac{1}{16}(x+1)^3 (3-x),
\end{equation}
and $\mathbb{K}(m)$ is an elliptic integral of the first kind,
\begin{equation}
\mathbb{K} (m) = \int_{0}^{1} \rd x \left[(1-x^2) (1 - m x^2)\right]^{-1/2}.\label{Elliptic-K-Wolfram}
\end{equation}

We stress that the definition~\eqref{Elliptic-K-Wolfram} corresponds to the notations of Wolfram Mathematica \cite{Wolfram}. The definitions of the complete elliptic integrals, for example, in \cite{Gradshteyn.book,Bateman.book,Byrd.book} employ the parameter $k^2$ as argument in place of the modulus $m$, viz. $\mathbf{K} (k) = \mathbb{K} (k^2)$. The purpose of the present brief report is to propose a more compact form of the DOS.

\section{Derivation}
For completeness, we recapitulate the main steps of the derivation that lead both to the new and old expressions for the DOS.
We begin with the tight-binding dispersion law of graphene written in the nearest-neighbor approximation \cite{Wallace1947PRev}
\begin{equation}
\epsilon(\mathbf{k})=  \pm t \sqrt{1+ 4 \cos^2
\frac{k_x a}{2} +4 \cos \frac{k_x a}{2} \cos \frac{\sqrt{3} k_y
a}{2}}\,,\label{dispersion}
\end{equation}
where  $\mathbf{k} = (k_x,k_y)$ is the wave-vector and $a = \sqrt{3} a_{\text{CC}}$ is the lattice constant
with $a_{\text{CC}}$ being the distance between the neighboring carbon atoms.

The DOS can be calculated as a trace of the imaginary part of the corresponding Green's function~\cite{Horiguchi1972JMP}
\begin{equation}
D(E) = - \frac{2 E}{\piup} \mbox{Im}[ g (E)],\label{DOS-def}
\end{equation}
with
\begin{equation}
g (E) = S \int_{BZ} \frac{\rd^2 k}{(2 \piup)^2} \frac{1}{(E+\ii 0)^2 - \epsilon^2(\mathbf{k})}\,.\label{G-tilde}
\end{equation}
The factor of $2$ in equation~\eqref{DOS-def} originates from the trace over the sublattice degree of freedom
and $S = \sqrt{3} a^2 /2$ in equation~\eqref{G-tilde} is the area of a unit cell. The integration is done over the
Brillouin zone (our notations correspond to \cite{Gusynin2007IJMPB}).
Introducing dimensionless variables and doubling the domain of integration to make it rectangular,
$-2 \piup/a \leqslant k_x \leqslant  2\piup/a $ and $-2 \piup /(a\sqrt{3}) \leqslant k_y  \leqslant 2\piup /(a\sqrt{3})$, one obtains
\begin{equation}
\! \! \!  g (E) = \frac{1}{8 \piup^2 t^2} \int \limits_{- \piup}^{\piup} \rd x  \int \limits_{- \piup}^{\piup} \rd y
\frac{1}{\tau - \cos 2x - 2 \cos x \cos y}
\end{equation}
with $\tau = (\varepsilon + \ri 0)^2/2 - 3/2$. Replacing $ s = \tan y/2$ we integrate over $s$ and obtain
\begin{equation}
g(E) = \frac{1}{4\piup t^2}\int \limits_{-\piup}^{\piup} \frac{\rd x}{\sqrt{\big(\tau - \cos 2x\big)^{2} - 4\cos^{2}x}}\,.\label{G-tilde-1st}
\end{equation}
Equation~\eqref{G-tilde-1st} can be expressed in terms of an elliptic integral of the first kind (see equation~(3.147.3) in~\cite{Gradshteyn.book})
\begin{equation}
g(E) = \frac{2}{\piup t^2 \sqrt{(\varepsilon-1)^{3}(\varepsilon+3)}}
\mathbf{K}\left(\sqrt{\frac{16\varepsilon}{(\varepsilon-1)^{3}(\varepsilon+3)}}\right),\label{G-tilde-2nd}
\end{equation}
where for the brevity of notations we omitted $+\ii 0$ in the argument.
The argument of the elliptic function in equation~\eqref{G-tilde-2nd} is imaginary for $0 \leqslant \varepsilon < 1$, but it is real and larger than 1  for $1 < \varepsilon \leqslant 3$ .
In the former case, $0 \leqslant \varepsilon < 1$, using the imaginary modulus transformation \cite{Byrd.book}
\begin{equation}
\mathbf{K}(\ii k) = \frac{1}{\sqrt{k^2 +1}}\mathbf{K}\left( \sqrt{\frac{k^{2}}{k^{2}+1}} \right)\label{imaginary}
\end{equation}
we arrive at the fist line of equation~\eqref{DOS-conventional}.
In the latter case, we use the relationship (see equation~(8.128) in \cite{Gradshteyn.book} and \cite{Byrd.book})
\begin{equation}
\mathbf{K}(k) = \frac{1}{k} \left[\mathbf{K} \left( \frac{1}{k}\right) - \ii \mathbf{K} \left( \sqrt{1 - \frac{1}{k^2}}\right) \right],
\qquad k >1,\label{k>1}
\end{equation}
where the sign in  front of the imaginary term is chosen in accordance to imaginary shift $\varepsilon + \ii 0$. Then, the last term of equation~\eqref{k>1}
leads us to the second line of equation~\eqref{DOS-conventional}.

\begin{figure}[!t]
\centering
\includegraphics[width=0.45\textwidth]{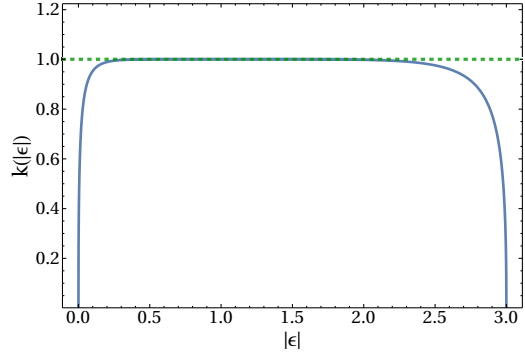}
\caption{(Color online) On the plot there is shown a dependence of the argument $k(|\varepsilon|) = \frac{4\sqrt{|\varepsilon|F(|\varepsilon|)}}{[\sqrt{|\varepsilon|}+\sqrt{F(|\varepsilon|)}]^{2}}$ of the elliptic integral in the DOS expression \eqref{DOS-final} on the modulus $|\varepsilon| = \frac{|E|}{t}$ (\emph{prepared with SciDraw}~\cite{SciDraw2005}).}
\label{fig:1}
\end{figure}

Using Landen's transformation (see equation~(8.126.3) in \cite{Gradshteyn.book,Byrd.book} and \cite{Gamayun2009PRB})
\begin{equation}
\frac{\theta(q-k)}{q} \mathbf{K} \left(\frac{k}{q} \right) + \frac{\theta(k-q)}{k} \mathbf{K} \left(\frac{q}{k} \right)
= \frac{1}{q+k} \mathbf{K} \left(\frac{2 \sqrt{qk}}{q+k} \right),\label{landen}
\end{equation}
the DOS~\eqref{DOS-conventional} can be represented in one line expression.
Returning back to the Wolfram's definition of the elliptic integral~\eqref{Elliptic-K-Wolfram}, we arrive at the final result of this report:
\begin{equation}
D(\epsilon) = \frac{1}{t \piup^2} \frac{|\varepsilon| \theta(3- |\varepsilon| )}{\sqrt{|\varepsilon|} + \sqrt{F(|\varepsilon|)}}
\mathbb{K} \left(\frac{4 \sqrt{| \varepsilon| F(|\varepsilon|)} }{\big[\sqrt{|\varepsilon|} + \sqrt{F(|\varepsilon|)}\big]^2 } \right).\label{DOS-final}
\end{equation}
Here, the argument of the elliptic integral is $\leqslant 1$ (see figure~\ref{fig:1}).

The unified result~\eqref{DOS-final} can be made clearer in the following way. By using~\eqref{G-tilde-2nd} and~\eqref{imaginary}, the expression~\eqref{DOS-def} for the DOS can be converted to the form

\begin{equation}
D(\epsilon) =\frac{4|\varepsilon|/(t\piup^{2})}{\sqrt{(|\varepsilon| + 1)^{3}(3-|\varepsilon|)}}\Re \mathbf{K}\left( \sqrt{\frac{16|\varepsilon|}{(|\varepsilon|+1)^{3}(3-|\varepsilon|)}}\right),
\end{equation}
valid for $0<|\varepsilon|<3$. The result~\eqref{DOS-final} then straightforwardly follows from analytical properties of the Green function; this is due to the identity
\begin{equation}
\Re \mathbf{K}(z) = \frac{1}{1+z}\mathbf{K}\left( \frac{2\sqrt{z}}{z+1}\right), \qquad 0<z<\infty
\end{equation}
following from~\eqref{landen}.

\section{Conclusions}
To conclude, we reproduce limiting cases of the DOS.
The low-energy expansion of the DOS~\eqref{DOS-final}~is
\begin{equation}
    D(\varepsilon) = \frac{1}{t} \left[\frac{2|\varepsilon|}{\sqrt{3} \piup}  +
    \frac{2 |\varepsilon|^3 }{3 \sqrt{3} \piup} + O(|\varepsilon|^5)  \right],\label{low-energy}
\end{equation}
where the first term originates from the contribution of $\mathbb{K} (0) = \piup/2$.
One can see that the second term of the expansion is 100 times smaller than the first one for
$|E| \lesssim  0.17 t \sim 0.5$~eV. Assuming that the Fermi velocity is $v_{\text F} = \sqrt{3} t a/(2 \hbar)$,
we reproduce the commonly used DOS per unit area and spin
$D(E) = |E|/(\piup \hbar^2 v_{\text F}^2) $. Finally, near $\varepsilon = 1$, the argument of the elliptic integral
in equation~\eqref{DOS-final} is $1 - (\varepsilon -1)^6/256$. Assuming that $\mathbb{K}(z) = -1/2 \ln(1-z)/16$ for $z \to 1$ \cite{Wolfram},
we reproduce the asymptotic of the DOS near the van Hove singularity \cite{Hobson1953PRev}
$D(E) = - 3 /(2\piup^2 t) \ln ( |1 - \varepsilon|/4)$. \\
\section*{Acknowledgements}

We thank S.G.~Sharapov for suggesting to reproduce the result of \cite{Hobson1953PRev} in his lecture course on graphene and for providing a great help in writing the article. Also, we would like to thank O.O. Sobol for a helpful and productive discussion.

\ukrainianpart

\title{Про вираз для густини станів у графені}
\author{В.О. Ананьєв, М.Ю. Овчинніков}
\address{Фізичний факультет, Київський національний університет імені Тараса Шевченка,\\ просп. Академіка Глушкова, 6, 03680 Київ, Україна}

\makeukrtitle

\begin{abstract}
\tolerance=3000%
Ми пропонуємо альтернативний аналітичний вираз для густини електронних станів у чистому графені в наближенні найближчих сусідів. На противагу вже відомим виразам, він представляє собою єдину формулу, справедливу на всьому інтервалі енергій. Також було перевірено відповідність уже відомим виразам і досліджено граничні випадки.
\keywords графен, густина станів, сингулярність Ван Гове
\end{abstract}
\end{document}